\documentstyle[12pt]{article}
\begin{document}
\newcommand{\nd}[1]{/\hspace{-0.5em} #1}
\begin{titlepage}
\begin{flushright}
hep-ph/9406406\\
June 25, 1994  \\
\end{flushright}

\begin{centering}
\vspace{.6in}
{\Large {\bf Solvability, Consistency and the Renormalization Group in
Large-$N_c$ Models of Hadrons}}

\vspace{.5in}
{\bf Nicholas Dorey }\\
\vspace{.1in}
Physics Department, University College of Swansea \\
Swansea, SA2 8PP, UK.$\quad$ pydorey@pygmy.swan.ac.uk \\
\vspace{.5in}
{\bf James Hughes} \\
\vspace{.1in}
Physics Department, Michigan State University \\
East Lansing, MI 48823, USA.$\quad$ hughes@msupa.pa.msu.edu \\
\vspace{.5in}
{\bf Michael P. Mattis} \\
\vspace{.1in}
Theoretical Division,  Los Alamos National Laboratory \\
Los Alamos, NM 87545, USA.$\quad$ mattis@skyrmion.lanl.gov \\
\vspace{.8in}
{\bf Abstract} \\
\vspace{.05in}
\end{centering}
{\small We establish the following fundamentals about
Lagrangian models of meson-baryon interactions in the large-$N_{c}$
limit: {\bf 1.} Calculating the leading-order contribution to
$1$-meson/$2$-baryon
Green's functions in the $1/N_c$ expansion involves summing an
infinite class of divergent Feynman diagrams. So long as the bare
Lagrangian properly obeys all
large-$N_c$ selection rules, this all-loops resummation is
accomplished exactly by solving coupled classical field equations with a
short-distance cutoff. {\bf 2.} The only effect of the resummation is
to renormalize the bare Yukawa couplings, baryon masses and hyperfine
baryon mass splittings of the model.  {\bf 3.}
In the process, the large-$N_{c}$ renormalization group flow of these
bare parameters is completely determined.  We conjecture that
 variants of the Skyrme model emerge as UV fixed points of such flows.}
\end{titlepage}
\section*{}
\paragraph{}
{\bf Introduction.} In the two decades since 't Hooft's original
suggestion \cite{GT}, little progress has been made towards a solution
of QCD in the limit $N_c\rightarrow\infty$. However,
several powerful constraints
on the allowed spectrum of baryons and their interactions with mesons
in this limit are known. These ``large-$N_c$
selection rules'' can be derived either from
the planar graphs in the underlying
quark-gluon theory \cite{EW,LM,Georgi}, from the Skyrme model
\cite{ANW,MatMuk},
from the nonrelativistic quark model \cite{M,MatBraat}, or finally
from the self-consistency of meson-baryon Feynman diagrams
\cite{GS,MPM,DM,J1}.
\paragraph{}
In this Letter, we
show how all these constraints can be incorporated consistently into an
effective hadron Lagrangian. Although
meson self-interactions become weak as $N_c\rightarrow\infty$,
meson-baryon Yukawa couplings become strong, and
an infinite number of diagrams dress the bare
Yukawa vertices of the theory at \it leading \rm
order in $1/N_c$. Our first result, extending Ref.~\cite{AM},
is that these diagrams can be summed exactly by solving coupled
classical field equations in the presence of a nonrelativistic
baryonic source; i.e., these models are {\em
solvable} in the $N_c\rightarrow\infty$ limit. Moreover, the only
effect of this resummation is to renormalize the bare baryon
masses, hyperfine mass splittings, and on-shell Yukawa  couplings of the
 theory, in such a way that the large-$N_c$ constraints
incorporated in the bare Lagrangian emerge unscathed at the
renormalized level. This confirms the
{\em consistency} of effective hadron Lagrangians at
large-$N_c$. Finally, the all-loops
{\em Renormalization Group (RG) equations}  controlling the
running of these bare quantities with the UV cutoff $\Lambda$
falls out as a by-product of the above analysis. The fact that classical
equations always suffice to leading order is a reflection of the important
fact that $N_c$ appears in the action in the combination $\hbar/N_c.$
\paragraph{}
The resulting picture of the large-$N_c$ baryon\footnote{This picture,
in which  a hard core of explicit nucleon degrees of freedom is dressed
by tree graphs (Born graphs) of mesons, is sometimes referred to
as ``hard-core Born-ography.''}
is reminiscent of the old chiral
bag models, in which explicit nucleon/quark degrees of freedom for
$r\le\Lambda^{-1}$ are matched onto a cloud of hedgehog pions at
$r\ge\Lambda^{-1}$ \cite{cloudy}.
 But whereas historically these bag models were motivated by
chiral symmetry, our current construction follows from large-$N_c$
\it alone\rm---with
no reference whatsoever to approximate chiral invariance. And since at
large distances our meson-dressed
baryon is indistinguishable from a skyrmion, it is natural
to conjecture variations of the Skyrme model as arising as UV fixed points
of the RG flow as $\Lambda\rightarrow\infty$.
\paragraph{}
{\bf Large-$N_c$ hadron models.} We study generic 2-flavor relativistic
hadron Lagrangians that conserve $C,$ $P,$ $T,$ and isospin,
and are further restricted \it only \rm by these
five large-$N_c$ consistency conditions:
\begin{description}
\item[1:] Straightforward quark-gluon counting arguments
show that $n$-meson vertices $\sim$  $N_c^{1-\frac{n}{2}}$,
 as do $n$-meson $2$-baryon vertices \cite{EW,LM,Georgi}.
Thus, baryon masses ($n=0$) and Yukawa couplings ($n=1$) grow like $N_c$
and $\sqrt{N_c}$, respectively.
\item[2:] The 2-flavor baryon spectrum of large-$N_c$ QCD consists of
an infinite tower of positive parity states with $I=J=1/2$, $3/2$,
$5/2\ldots$. To leading order
 these states are degenerate, with mass $M_{\rm bare}\sim N_{c}$
\cite{LM,Georgi,ANW,GS,DM}.
\item[3:] Hyperfine baryon mass splittings
 have the form $J(J+1)/2{\cal I}_{\rm bare}$ where ${\cal I}_{\rm bare}
\sim N_c$ \cite{LM,Georgi,ANW,J1}.
\item[4:] Yukawa couplings are constrained to obey
the ``proportionality rule'' \cite{ANW,MatMuk,MatBraat,GS,DM}, which
fixes the interaction strength of
a given meson with each member of the baryon tower as a multiple of one
overall coupling constant (e.g., $g_{\pi N\Delta}/g_{\pi NN}=3/2$).
\item[5:] Finally, the allowed couplings of mesons to the baryon tower
must obey the $I_{t}=J_{t}$ rule \cite{Georgi,MatMuk,MatBraat,MPM}; e.g.,
the $\rho$ meson must be tensor-coupled to the nucleon
while the $\omega$ meson is vector-coupled at leading order in
$1/N_{c}$, in good agreement with phenomenology.
\end{description}
\paragraph{}
{\bf Summing the leading-order graphs.} The $N_c$ dependence of
coupling constants described in {\bf 1} above suffices to identify
the leading-order meson-baryon Feynman graphs for any given physical process.
Thus, purely mesonic processes are dominated by meson
tree graphs, which vanish as $N_c\rightarrow\infty$.
Correspondingly,
meson-baryon processes are dominated by those graphs which become
meson trees if the baryon lines are removed.
To illustrate the complexity of such graphs,  look at a typical multi-loop
correction, Fig.~1b, to the bare Yukawa coupling, Fig.~1a. Since,
by design, the graph in Fig.~1b contains no loops formed purely from
mesonic legs, this graph scales like $\sqrt{N_c}$ just like the bare
vertex. This is trivially checked by multiplying together all the
vertex constants and ignoring propagators entirely, since both meson
and baryon propagators $\sim N_{c}^{0}$. Hence,
in order to calculate the dressed Yukawa vertex to leading
order, one must sum this infinite set of diagrams. One must \it also \rm
sum all multiple
insertions of the baryon self-energy corrections and additional vertex
corrections as illustrated in Fig.~1d, as these too contribute at leading
order \cite{LM,J1}.
 Since many of the loop integrations in  these diagrams are UV
divergent, it is necessary to regulate the theory with a UV cutoff $\Lambda$.
\paragraph{}
Following the methods of Ref.~\cite{AM}, a summation of \it all \rm
such graphs is possible, to leading order in $1/N_c.$ To see how,
we  focus initially on a  model of
baryons and pions only, in which the above five conditions are imposed:
\def\vpi{\vec\pi}
\begin{eqnarray}
{\cal L} & = & \frac{1}{2}(\partial_{\mu}\vpi)^2
-\frac{m_{\pi}^2}{2}\vpi^2 - V(\vpi)+
\bar{N}(i\nd{\partial}-M_{N})N \nonumber \\ & & \qquad{}
-g^{\rm bare}_{\pi}\partial_{\mu}\pi^{a}\bar{N}\gamma^{5}
\gamma^{\mu}\tau^{a}N + \hbox{(higher-spin baryons)}
\label{pionly}
\end{eqnarray}
The incorporation of additional meson species  will be discussed below.
Here $V$ is a general pion potential including quartic and higher vertices,
and $M_{N}=M_{\rm bare}+3/8{\cal I}_{\rm bare}$
including the hyperfine splitting.
The pseudovector form of the $\pi N$ coupling is determined by the
$I_{t}=J_{t}$ rule while the proportionality rule fixes the
corresponding pion couplings to the higher-spin baryons.
\paragraph{}
{}From the position-space Feynman rules for (\ref{pionly}),
the sum of all the graphs such as Fig.~1b is formally given by:
\def\calY{{\cal Y}}
\def\calA{{\cal A}}
\begin{equation}
\sum_{n=1}^{\infty}\int \left(\prod_{i=1}^{n} d^4w_{i}d^4z_{i}\right)
Blob_{n}(x;w_{1}, \ldots,w_{n}) \sum_{\rho \in S_{n}}\prod_{i=1}^{n}
g(w_{i},z_{\rho(i)})\cdot \calY(z_{\rho(i)})\prod_{i=1}^{n-1} G(z_{i},z_{i+1})
\label{formal}
\end{equation}
$Blob_{n}$ denotes the shaded blob in Fig.~1c; the sum over
permutations $\rho \in S_{n}$ counts the $n!$ possible tanglings of the $n$
meson lines which connect to the blob. All isospin and
spin indices have been suppressed in (\ref{formal});
$g_{ab}$ and $G_{J}$ are the
position-space meson  and spin-$J$ baryon propagators, respectively, and
$\calY_{JJ'}^a$ is the appropriate pseudovector Yukawa vertex factor.
\paragraph{}
Passing to the large-$N_c$ limit, we can exploit two important simplifications
to this expression. First, the baryons become very massive and can be
treated nonrelativistically. For forward time-ordering, $z_{0}<z'_{0}$, the
baryon propagator $G(z,z')$ can be replaced by its
nonrelativistic counterpart $G_{NR}(z,z')+{\cal O}(1/N_{c})$. As usual,
the reversed or ``$Z$-graph'' time ordering $z'_{0}<z_{0}$
contributes effective pointlike vertices in which two or more mesons couple to
the baryon at a single point. These effective ``seagull'' terms are naturally
grouped with similar pointlike interactions that one may wish to incorporate
from the outset in the bare Lagrangian, and
we will mention below the simple generalization of
our analysis required to account for both these sets of terms consistently.
\paragraph{}
A second simplification comes from switching to the
$SU(2)$ ``collective coordinate basis'' $|A\rangle$ for the $I=J$ baryon tower
\cite{ANW,M,MatBraat},
 related to the more familiar spin-isospin basis
$|I=J,i_{z},s_{z}\rangle$ via
\begin{equation}
\langle I=J, i_{z}, s_{z}|A\rangle = ({2J+1})^{1/2}D_{-s_{z},i_{z}}^{(J)}
(A^\dagger)\cdot(-)^{J-s_{z}}
\end{equation}
with $D^{(J)}(A)$  a  Wigner $D$-matrix.
In the $|A\rangle$ basis, the baryon propagator can be expressed as a \it
quantum mechanical \rm path integral over two collective coordinates:
${\bf X}$,  representing the position of the center of the baryon, and $A$,
describing its $SU(2)$ (iso)orientation,
\begin{eqnarray}
G_{NR}^{A,A'}(z,z') & = & \theta(z'_{0}-z^{}_{0})
\int_{{\bf X}(z_{0})={\bf z}}
^{{\bf X}(z'_{0})={\bf z}'}{\cal D}{\bf X}(t)\int_{A(z_{0})=A}
^{A(z'_{0})=A'}{\cal D}A(t) \nonumber
\\
& &  \qquad{} \times \exp{\left(i\int_{z_{0}}^{z'_{0}}dt\,
M_{\rm bare}+\frac{1}{2}M_{\rm bare}\dot{{\bf X}}^{2}
+{\cal I}_{\rm bare}{\rm Tr}\dot{A}^{\dagger}\dot{A}\right)}
\label{gnr}
\end{eqnarray}
The path integration over $A(t)$ can be performed using the beautiful
result of Schulman for free motion on the $SU(2)$ group manifold
\cite{schul},
\begin{eqnarray}
& &\int_{A(z_{0})=A}^{A(z'_{0})=A'}{\cal D}A(t)\,
\exp i\int_{z^{}_{0}}^{z'_{0}}dt\,
{\cal I}_{\rm bare}{\rm Tr}\dot{A}^{\dagger}\dot{A}\quad=
\nonumber
\\
& &\sum_{J=1/2,\,3/2,\cdots}\ \sum_{i_z,s_z=-J}^J
\langle A'|J,i_z,s_z\rangle\cdot
\exp i(z_0'-z^{}_0){J(J+1)\over2{\cal I}_{\rm bare}}\cdot
\langle J, i_{z}, s_{z}|A\rangle\ ,
\label{schu}
\end{eqnarray}
yielding the conventional nonrelativistic propagator for an infinite tower of
particles with masses $M_{\rm bare}(J)=M_{\rm bare}+
J(J+1)/2{\cal I}_{\rm bare}$ as required.
\paragraph{}
The greatest advantage of the $|A\rangle$ basis is
that the Yukawa vertex factor $\calY$ for the pion-baryon
coupling becomes \it diagonal \rm \cite{ANW,M,MatBraat}:
\begin{equation}
\calY_{A,A'}^{a}(z)=-3g^{\rm bare}_{\pi}D_{ab}^{(J)}(A)\delta(A-A')
\frac{\partial}{\partial z_{b}}
\end{equation}
Substituting for $G$ and $\calY$ in (\ref{formal}), we find that
this diagonality property allows us trivially
to perform the sum over the $n!$ products
of step functions, which collapses to unity.
Interchanging the order of path integration and the product over baryon
legs, one obtains
\begin{eqnarray}
& &\int {\cal D}{\bf X}(t){\cal D}A(t)\sum_{n=1}^{\infty}
 \int \left( \prod_{i=1}^{n} d^4w_{i}d^4z_{i}\right)
Blob_{n}(x;w_{1}, \ldots,w_{n})
\nonumber
\\
& &\times\ \prod_{i=1}^{n}
g(w_{i},z_{i})\cdot \calY(z_{i}) \delta^{(3)}_{\Lambda}
\big({\bf z}_{i}-{\bf X}(z^{0}_{i})\big)
\exp i S_{\rm baryon}[{\bf X},A]
\label{formal2}
\end{eqnarray}
where $S_{\rm baryon}$ is short for
the exponent of (\ref{gnr}). We have  reduced the problem to a
sum of \it tree \rm diagrams for the pions interacting with the baryon
collective coordinates through a $\delta$-function source. The only remaining
manifestation of the UV cutoff $\Lambda$ is that this $\delta$-function
should be smeared out over a radius $\sim \Lambda^{-1}$, as
 denoted by $\delta_{\Lambda}$ in (\ref{formal2}), which we assume still
preserves rotational invariance.
\paragraph{}
In short, the massive baryon has become a translating, (iso)rotating,
smeared point-source for the
pion field, the effect of which can be found be solving the
appropriate \it classical \rm Euler-Lagrange equation for a configuration
we call $\vec{\pi}_{\rm cl}(x;[{\bf X}],[A])$ \cite{AM}:
\def\sqr#1#2{{\vcenter{\vbox{\hrule height.#2pt
        \hbox{\vrule width.#2pt height#1pt \kern#1pt
           \vrule width.#2pt}
        \hrule height.#2pt}}}}
\def\square{\mathchoice\sqr84\sqr84\sqr53\sqr43}
\begin{equation}
(\square+m_{\pi}^{2})\pi_{\rm cl}^{a}+\frac{\partial
V}{\partial
\pi_{\rm cl}^{a}}=3g^{\rm bare}_{\pi}D_{ai}^{(1)}\big(A(t)\big)
\frac{\partial}{\partial x^{i}}
\delta^{(3)}_{\Lambda}({\bf x}-{\bf X}(t))
\label{ceq}
\end{equation}
It is easily checked (Fig.~2) that the order-by-order
perturbative solution of Eq.~(\ref{ceq}) generates precisely the sum of
graphs appearing in (\ref{formal2}).
By similar semi-classical reasoning (Fig.~3), the
leading-order parts of
the additional vertex and self-energy corrections highlighted in Fig.~1d
exponentiate exactly, and are correctly accounted for by evaluating
the mesonic plus Yukawa pieces of the action (call this sum $S_{\rm eff}$)
on $\vpi_{\rm cl}$.
The final leading-order
result for the complete sum of graphs contributing to the dressed
pion-baryon vertex is
\begin{equation}
\int {\cal D}{\bf X}(t){\cal D}A(t)\, \pi^{a}_{\rm cl}(x;[{\bf X}],[A])
\exp i \big(
S_{\rm baryon}+ S_{\rm eff}[\vec{\pi}_{\rm cl},{\bf X}(t), A(t)]\,\big)
\label{ans}
\end{equation}
\paragraph{}
{\bf Solving the classical field equation.}
We solve Eq.~(\ref{ceq}) by relating it to the analogous equation
for the \it static \rm pion cloud, $\vec{\pi}_{\rm stat}({\bf x})$,
surrounding a fixed baryon source (${\bf X}(t)\equiv{\bf0}, A(t)\equiv 1$):
\begin{equation}
(-\nabla^{2}+m_{\pi}^{2})\pi_{\rm stat}^{a}+\frac{\partial
V}{\partial\pi_{\rm stat}^{a}}=3g^{\rm bare}_{\pi}\frac{\partial}{
\partial x^{a}}\delta^{(3)}_{\Lambda}({\bf x})
\label{ceq2}
\end{equation}
The solution will generically have the hedgehog form
familiar from the Skyrme model:
$\pi^{a}_{\rm stat}({\bf x})=(f_{\pi}x^{a}/2r)F(r)$ where $r=|{\bf x}|$. The
profile function $F(r)$ is found, in turn,
by solving the induced nonlinear radial ODE.
While its detailed form depends sensitively on
the potential $V(\vpi)$,  its asymptotic behavior for
large $r$ is fixed by the linearized field equation,
\begin{eqnarray}
F(r) & \longrightarrow & \calA\left(\frac{m_{\pi}}{r}+\frac{1}{r^{2}}\right)
e^{-m_{\pi}r} \
\label{asm}
\end{eqnarray}
where the constant $\calA$ must be extracted numerically.
The solution to  (\ref{ceq}) is then simply given, up
to $1/N_{c}$ corrections, by translating and (iso)rotating $\vpi_{\rm stat}$:
\begin{equation}
\pi_{\rm cl}^{a}(x;[{\bf X}],[A])=D^{(1)}_{ab}\big(A(t)\big)\pi_{\rm stat}^{b}
\big({\bf x}-{\bf X}(t)\big)
\end{equation}
The additional collective coordinate dependence carried by $\vpi_{\rm cl}$
versus $\vpi_{\rm stat}$ is precisely that required for overall isospin,
angular momentum and 4-momentum conservation, as is easily checked
\cite{DHM}.
\paragraph{}
We seek the renormalized on-shell $\pi N$ interaction, to leading
order in $1/N_c.$ It is defined in the usual way
as the on-shell residue of the LSZ amputation of the full
set of graphs that are summed implicitly by
 Eq.~(\ref{ans}). Formally, this amputation is
identical to the procedure one follows
in the Skyrme model \cite{DHM}. In particular, the
physically correct analytic structure of the one-point function
follows from the  $1/N_c$ corrections to $\vpi_{\rm cl}$ which describe
its response to the rotation of the source. (The specifics of this
response, involving an interesting small distortion \it away \rm
from the hedgehog ansatz \cite{DHM}, need not
concern us here.) Thanks to the (iso)vector nature of the hedgehog,
the resulting S-matrix element for one-pion
 emission defines a renormalized on-shell pseudovector interaction of the pion
with the baryon tower, {\em identical} to the bare interaction in
 (\ref{pionly}), except for the coupling constant renormalization
 $g_{\pi}^{\rm bare}\rightarrow g_{\pi}^{\rm ren}.$ Again as in the
Skyrme model, this latter quantity is
determined by the asymptotics of $\vec{\pi}_{\rm stat}$, Eq.~(\ref{asm}),
and is explicitly given by \cite{ANW,DHM}
$g_{\pi}^{\rm ren}=(2/3)\pi f_{\pi}\calA$. Thus the proportionality and
$I_{t}=J_{t}$ rules for the pion-baryon coupling remain
true at the renormalized level, as claimed.
\paragraph{}
Furthermore, the result of evaluating $S_{\rm eff}[\vpi_{\rm cl}]$
is just an additive renormalization of the bare parameters of $S_{\rm baryon}$,
due to the meson cloud:
\begin{equation}
S_{\rm baryon}+S_{\rm eff}[\vec{\pi}_{\rm cl},{\bf X},A]\ =  \
\int dt\,\Big(
M_{\rm ren}+\frac{1}{2}M_{\rm ren}\dot{{\bf X}}^{2}
+{\cal I}_{\rm ren}{\rm Tr}\dot{A}^{\dagger}\dot{A}\,\Big)
\end{equation}
where
\begin{eqnarray}
M_{\rm ren}=M_{\rm bare}+\int d^{3}{\bf x}\,
({\nabla}\pi_{\rm cl}^{a})^2\ ,
& & {\cal I}_{\rm ren}={\cal I}_{\rm bare}+\frac{2}{3}\int d^{3}{\bf x}\,
\vpi_{\rm cl}^2
\end{eqnarray}
It follows that
$M_{\rm ren}(J)=M_{\rm ren}+J(J+1)/2{\cal I}_{\rm ren}$ and so the form
of the hyperfine mass splitting is likewise preserved by renormalization.
\paragraph{}
The generalization of the above analysis to models including several
species of mesons  involves solving the
 coupled classical radial ODE's for all the meson fields, using
generalized hedgehog ansatze familiar from
vector-meson-augmented Skyrme models.
A particularly rich meson model might include, in addition to
the pion, the tensor-coupled
$\rho$, i.e., $g_\rho^{\rm bare}
\partial_\mu\vec\rho_\nu\cdot\bar N\sigma^{\mu\nu}
\vec\tau N,$ the vector-coupled $\omega,$ i.e.,
$g^{\rm bare}_\omega
\omega_\mu\bar N\gamma^\mu N,$ and/or the ``$\sigma$-meson,''
which couples simply as $g_\sigma^{\rm bare}
\sigma \bar N N.$ Again, on shell, the form
of these particular couplings survives renormalization.
Multi-meson  ``sea-gulls,'' defined earlier, can
also be incorporated. For $\Lambda$ sufficiently large, these have the
effect of multiplying the right-hand side of Eq.~(\ref{ceq}) by a
function of the field, which simply leads to  further
\it algebraic \rm renormalizations of the bare Yukawa constants.
\paragraph{}
{\bf Large-$N_{c}$ Renormalization Group.} We have
described an  explicit numerical procedure
for calculating the renormalized Yukawa couplings, baryon
masses and hyperfine mass splittings,
to leading order in $1/N_{c}$, directly from the classical meson cloud
surrounding the baryon. Since the $\delta$-function
source on the right-hand side of Eq.~(\ref{ceq}) is smeared out over a
characteristic length $\Lambda^{-1}$, these
quantities depend explicitly on $\Lambda$. In order to hold the
physical, renormalized masses and couplings fixed,
 it is necessary to vary simultaneously
{\em both} $\Lambda$ and the corresponding bare quantities.
This procedure defines an RG flow for
$M_{\rm bare}(\Lambda)$, ${\cal I}_{\rm bare}
(\Lambda)$ and $g_{\pi,\rho,\omega,\sigma}^{\rm bare}(\Lambda)$,
valid to all orders in the loop expansion but strictly to leading order in
$1/N_{c}$. Numerical work along these lines is in progress.
\paragraph{}
It is particularly interesting to speculate whether these RG flows exhibit a
UV fixed point, meaning that the theory has a
 nontrivial continuum limit. Such a
limit---{\it if} it exists---must be defined with care.
 As the validity of nonrelativistic baryon propagators
breaks down when the momenta of the impinging mesons $\sim\,M_N$, it
is natural to restrict $\Lambda\ll M_N\sim N_c$; e.g.,
$\Lambda\sim\sqrt{N_c}.$ Basically, this is tantamount to only taking
the  continuum limit $\Lambda\rightarrow\infty$
\it after \rm letting $N_c\rightarrow
\infty,$ an important caveat, as it is likely that these limits do not commute.
Moreover, if $\Lambda\rightarrow\infty$ while the bare Yukawa
couplings stay finite, the nonlinearity of Eq.~(\ref{ceq}) ensures that it
no longer has a meaningful solution. This suggests that
the bare Yukawa couplings must flow to zero in this limit.
Consequently, the continuum theory is nontrivial if
the \it homogeneous \rm variant of Eq.~(\ref{ceq2}) (with the right-hand
side set to zero) admits a nonperturbative solution, i.e., a
soliton/skyrmion (either energetic or topological). We are left with
the conjectural but conceptually pleasing
 picture of variants of the Skyrme model emerging, in large-$N_c,$ as UV
fixed points of a more pedestrian class of models that have dressed
explicit fields  representing the nucleons, $\Delta$'s, etc. Note that
purely mesonic couplings and masses do \it not \rm flow in
our program, since meson loops $\sim 1/N_c.$ This suggests that we use
the experimental, renormalized meson parameters from the outset---implicitly
summing all such loops! Note that only for a meson parameter space
of measure zero can there be a UV fixed point of the type just described,
since, obviously, in Skyrme-type models the renormalized meson
parameters, Yukawa couplings, and baryon masses cannot be independently
fixed, but are tightly interconnected (which is precisely the point
of the Skyrme approach).
\paragraph{}
\bf Meson-baryon scattering. \rm The above formalism for dressed Yukawa
couplings is naturally extended to physical processes in which the baryon
interacts with more than one asymptotic meson. Figures 4a and 4b illustrate
all
the leading-order contributions, ``Compton-type'' versus ``exchange-type''
respectively, to meson-baryon scattering. After resummation/renormalization,
the
Compton-type graphs turn into the two elementary time-ordered graphs 5a and 5b,
where now the vertices are the \it renormalized \rm Yukawa couplings. The
exchange-type graphs sum to Fig.~5c, where the thick meson line stands
for the propagation of the fluctuating meson field, say $\delta\vpi$
(defined as $\vpi-\vpi_{\rm cl}$), through the nontrivial background
generated by $\vpi_{\rm cl}$ itself \cite{AM}. It is pleasing that, despite
appearances, Figs.~5a-c can be viewed in a unified semiclassical
manner, as we explained in Ref.~\cite{DHM} (Sec.~7).
\paragraph{}
We thank T. Bhattacharya, A. Kovner, M. Peskin and R. Silbar for
incisive critique. ND acknowledges the Nuffield
Foundation for financial support. Similar conclusions are
reached by A. Manohar (UCSD-PTH/94-14, to appear).

\section*{Figure Captions}
\paragraph{}
\bf1\rm.
(a) The bare meson-baryon coupling, which we shall refer to generically
as a ``Yukawa coupling.'' Henceforth, directed lines are baryons,
undirected lines are mesons. Internal baryon lines must be summed over
all allowed states in the $I=J$ tower.
(b) A typical multi-loop dressing of (a)
that contributes at \it leading \rm order, $N_c^{1/2},$ as it contains no
purely mesonic loops. (c) A systematic counting of the diagrams such
as (b). The shaded blob contains only tree-level meson branchings.
There are $n!$ distinct ``tanglings'' of the
attachments of the shaded blob to the baryon line. (d) A typical
dressing such as (b), augmented by additional baryon self-energy and
vertex corrections, all of which also contribute at leading order.
\paragraph{}
\bf2\rm.
The graphical perturbative solution to Eq.~(\ref{ceq}) as a sum of
tree-level one-point functions terminating in the effective Yukawa
vertex.
\paragraph{}
\bf3\rm.
Diagrammatic representation of $S_{\rm eff}(\vpi_{\rm cl}).$ When
combined with the expansion depicted in
Fig.~2, $\exp iS_{\rm eff}$ combinatorically correctly accounts for
all the leading-order baryon self-energy and vertex corrections
highlighted in Fig.~1d.
\paragraph{}
\bf4\rm.
Typical ``Compton-type'' (a) and ``exchange-type'' (b) leading-order
contributions to meson-baryon scattering. In an exchange-type graph,
one can trace a path from the incoming meson $\phi$ to the outgoing
meson $\phi'$ without ever traversing a baryon line segment; in
a Compton-type graph one cannot.
\paragraph{}
\bf5\rm.
The three leading-order \it renormalized \rm contributions to
meson-baryon scattering, which account for all the graphs of the
type illustrated in Fig.~4.

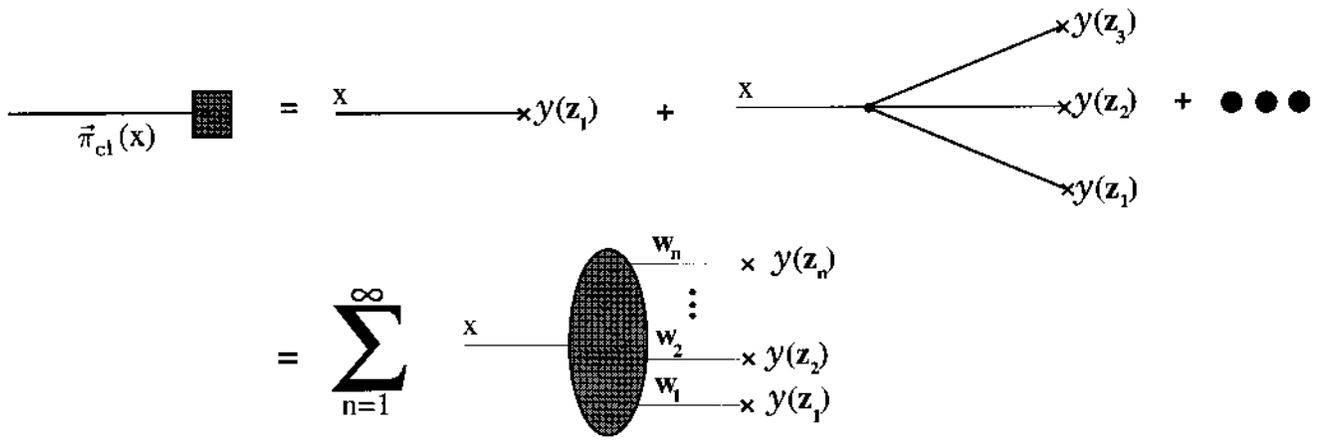
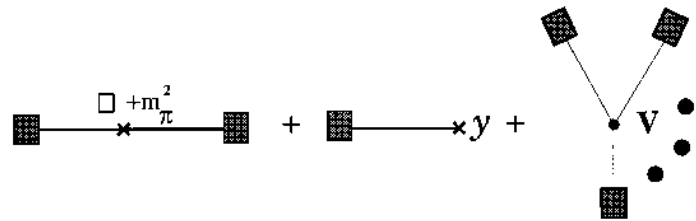
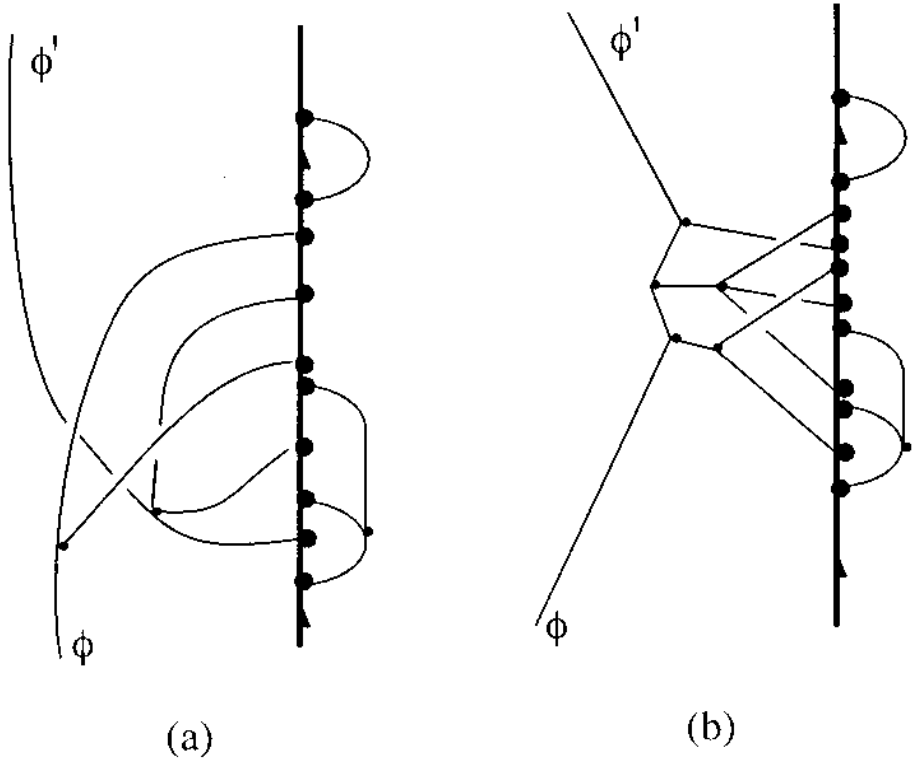
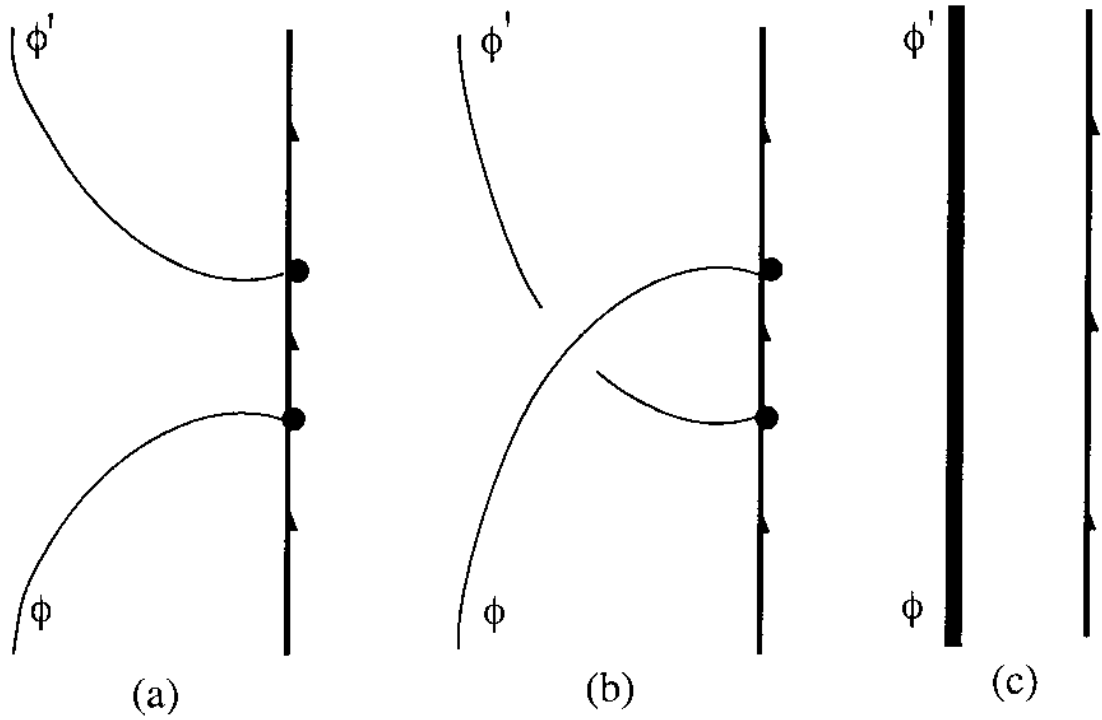
\end{document}